# Data-Driven Computation of the Accessibility Provided by Demand-Responsive Transport

**Pierfrancesco Leonardi**
Department of Civil Engineering and Architecture, University of Catania
Via Santa Sofia, 64, 95123 Catania, Italy
Email: pierfrancesco.leonardi@phd.unict.it

**Vincenza Torrisi**
Department of Electric, Electronic and Computer Engineering, University of Catania
Via Santa Sofia, 64, 95123 Catania, Italy
Email: vincenza.torrisi@unict.it

**Andrea Araldo**
Samovar, Télécom SudParis, Institut Polytechinque de Paris
19 place Marguerite Perey, 91120 Palaiseau, France
Email: andrea.araldo@telecom-sudparis.eu

**Matteo Ignaccolo**
Department of Civil Engineering and Architecture, University of Catania
Via Santa Sofia, 64, 95123 Catania, Italy
Email: matteo.ignaccolo@unict.it

## ABSTRACT

Conventional Public Transport (PT) cannot support the mobility needs in weak demand areas. Such areas could be better served by integrating, within PT, Demand-Responsive Transport (DRT), in which bus routes dynamically adapt to user demand. While the literature has focused on the level of service of DRT, it has overlooked its contribution to *accessibility*, which measures the ease of accessing opportunities (e.g, schools, jobs, other residents). Therefore, the following simple question remains unanswered: *How many additional opportunities per hour can be reached when DRT is deployed?*

However, no method exists to quantify the accessibility resulting from the integration of conventional PT and DRT. We propose a novel method to compute isochrone-based accessibility. The main challenge is that, while accessibility isochrones are computed on top of a graph-model of the transport system, no graph can model DRT, since its routes are dynamic and stochastic. To overcome this issue, we propose a data-driven method, based on the analysis of multiple days of DRT operation.

The methodology is tested on a case study in Acireale (Italy) simulated in Visum, where a many-to-many DRT service is integrated with a metropolitan mass transit. We show that, regarding the currently deployed conventional bus lines, the accessibility provided by DRT is much higher and its geographical distribution more equal. While current DRT planning is based exclusively on level of service and cost, our approach allows DRT planners and operators to shift their focus on accessibility.

**Keywords:** Accessibility; Multimodal Transport Network; Demand-Responsive Transport (DRT); Equity

**INTRODUCTION**

One main objective of transport planning is to ensure *accessibility*, i.e., the ability of people to reach opportunities (e.g., services, activities, jobs, other people) via the available transport options (1), (2), (3), (4). However, applying this concept *no matter* the transport options has resulted in car-oriented cities. Recognizing the related environmental, social and economic sustainability issues and in the effort to reduce externalities (5), planning strategies must now foster a modal shift from private vehicles to more sustainable modes, such as active mobility and Public Transport. Since active mobility can only cover a limited fraction of required accessibility (due to limits in trip lengths, city morphology, etc.), the role of Public Transport (PT) is of paramount importance. This motivates why we focus here on the *accessibility provided by PT*, i.e., the ability of people to reach opportunities *via PT*.

PT accessibility is unequally distributed in metropolitan areas (6) and is insufficient in the suburbs, which therefore suffer from a chronic car-dependency (7). Indeed, to prevent the PT cost per passenger from exploding in weak and scattered demand areas, PT operators can only provide low stop density and line frequencies, which results in a poor level of service and insufficient PT accessibility. This issue can be tackled via Demand Responsive Transport (DRT), where a fleet of buses is dynamically routed to better adapt to a weak demand (8). Integrating DRT with conventional PT has the potential to improve level of service (9), (10) and reduce the accessibility inequality commented above (11). In some limited areas, it must also be beneficial to replace conventional PT with DRT (12).

Generally, DRT requires additional cost of investment and operation (10). The reluctance of operators to bear this cost can be partially explained by the fact that, up to now, it has not been possible to quantify its benefits in convincing terms. Indeed, DRT has mainly been studied in terms of operational parameters and metrics (e.g., travel and waiting times, fleet dimensioning, etc.) that, by themselves, cannot justify the high additional cost. A much more convincing argument would be to show the improvement of accessibility, i.e., to understand "how many additional opportunities per hour can be reached thanks to DRT". However, no methods are available today to make this type of quantification. Assessing the accessibility provided by DRT is indeed challenging: accessibility measures are generally calculated on a graph. However, a DRT service cannot be represented as a graph, due to its stochasticity, as DRT vehicle trips change every day, depending on user requests.

In this paper, we remove this barrier by proposing a new method to assess the accessibility of DRT. The method is *data-driven*, as it is solely based on a simple statistical analysis of trips observed in previous days of DRT operations and does not require the development of complex and hard to validate econometric or simulation models. The key idea of the method is the description of DRT operation in terms of time-dependent graphs and the treatment of accessibility as a random variable, to account for the stochasticity of DRT. We showcase our method in a case study in Acireale (Italy), simulated in Visum, where we show the improvement of accessibility when replacing conventional PT with DRT, as well as a more equal geographical distribution of accessibility.

By allowing to assess the improvement of accessibility provided by DRT, we believe our work can contribute to a better understanding of this system, an increased acceptance by planners, operators and users, and a better willingness to undertake the economic investment needed for its deployment.

**RELATED WORK**

We focus in this section on the literature focused on the measurement of the impact of flexible mobility (shared taxis, DRT, etc.) on the accessibility. Accessibility measures based on random utility have been applied to on-demand mobility (13,14,15). The limit of such approach is that it requires the development of agent-based simulation models to measure the utilities perceived by individual users. The high complexity of developing the models, as well as calibrating and validating them, limits their scalability, replicability, ease of interpretation and trust that deciders can have in their results. These difficulties limit their application in practical cases.

Accessibility measures based on isochrones (16) are a valuable alternative to random utility accessibility measures, as they provide an easy-to-understand indicator, which basically counts the number of opportunities that can be reached in a given time interval. There is some relevant recent work computing isochrone-based accessibility for shared or flexible mobility. Considering the work proposed by Abouelela et al. [2024] (17), the authors compute an isochrone-based accessibility measure for a shared e-scooter service. The limitation is that (i) they cannot capture the fact that multiple trips are shared among users, which however determines the economic and environmental efficiency of DRT, (ii) they cannot capture the accessibility of flexible modes *combined with* conventional PT.

Le Hasif et al. [2022] (18), Wang et al. [2024] (19) adopt analytical models based on continuous approximation to roughly estimate DRT travel times, and compute accessibility based on estimations. While such approaches can be a useful starting point in strategic planning, to reason about where DRT deployment is more appropriate, their accessibility do not capture the real accessibility experienced by users, as analytical models are unsuited to do so.

Diepolder et al., 2023 (20) are the first to propose a method to compute isochrone-based accessibility based on empirical observations of DRT trips. The DRT considered in their work is *many-to-one*: DRT buses bring multiple passengers at a time to/from major rail hubs, providing a feeder service around them. Many-to-one DRT is appropriate in areas with a sufficient presence of conventional PT stops. This might not be the case of weak demand areas, where door-to-door many-to-many DRT services, providing a direct connection from users' origins to destinations, is more appropriate. The method of Diepolder et al. 2023 is based on modeling travel and waiting times as 2-dimensional random fields around each conventional PT stop. Their method relies on geostatistical analyses that exploit geographical locality: (i) the similarity of travel and waiting times of user trips within close-by origins, going to the same conventional PT stop and (ii) the similarity of travel times of user trips starting from the same conventional PT stop and going to close-by destinations. Therefore, their method cannot generalize to a many-to-many DRT service. Indeed, the user trips served in such a service cannot be positioned on a 2-dimensional surface (as we have to consider the (x,y) of the origin and the (x,y) of the destination – 4 dimensions in total), which prevents from using the 2-dimensional random field formalism and subsequent analysis. In this paper we overcome this limitation by a more generic approach, which can handle trips *whatever they are*, requiring neither any many-to-one structure nor any geographical locality.

To the best of our knowledge, we are the first to provide a method to quantify isochrone-based accessibility of a generic DRT service (many-to-one or many-to-many), which captures trip sharing and multimodality (i.e., combination with conventional PT), and which is solely based on previous observed DRT trips. In terms of implementation and data format, we point out the availability of the GTFS-Flex format. GTFS-Flex extends the General Transit Feed Specification (GTFS), used to describe conventional PT schedules, with new tables for the description of DRT services (21). In our implementation, we describe DRT in the original GTFS format, but we may consider using GTFS-Flex in the future.

## METHODOLOGY

### Preliminaries

*Accessibility*

We aim to compute the accessibility metrics by Biazzo et al. (22). We now explain their calculation in the case of conventional PT. First, the study area is zoned with a regular tessellation in hexagons. We remove the hexagons whose distance between the respective centroid and the closest PT stop is larger than *max_walk_distance*. We compute accessibility scores for the remaining hexagons. Two metrics are proposed, i.e. velocity score and the sociality score.

For a generic hexagon $i$ we firstly set a departure time $t_0$ then calculate area $A(t, i)$ that can potentially be visited when starting at the centroid of hexagon $i$ at time $t_0$, within time $t$, using walk and PT lines.

Transfers from line to line are possible and waiting times are considered whenever a line is boarded. We then make the simplifying assumption that area $A(t,i)$ is a circle and compute its radius:

$$r(t,i) = \sqrt{\frac{A(t,i)}{\pi}} \tag{1}$$

We then calculate a corresponding velocity $V(t,i) = \frac{r(t,i)}{t}$ To compute a velocity score of hexagon $i$ at time $t_0$, we need to get rid, in the previous formula, from the dependence on user trip duration $t$. To this aim, let $f(t)$ be the probability distribution of a user trip duration, which can be computed based on trip times usually observed in cities. The **velocity score** of hexagon $i$ at time $t_0$ is:

$$V(i) = \int_0^\infty V(t,i) \, f(t) \, dt \tag{2}$$

The calculated velocity score provides a measure of the average speed each individual experiences while traveling on public transport services, starting from a chosen hexagon going in any direction.

To compute an isochrone-based accessibility score, we first compute number $S(t,i)$ of opportunities reachable within time t when starting at instant $t_0$ from hexagon $i$. $S(t,i)$ corresponds to the opportunities located inside area $A(t,i)$. Similarly, as before, we compute the **accessibility score** as:

$$S(i) = \int_0^\infty S(t,i) \, f(t) \, dt \tag{3}$$

The accessibility score can be interpreted as how many opportunities can be reached in a typical user trip thanks to PT. For simplicity and availability of data, we consider, as in (22), people as opportunities, and call $S(i)$ *sociality score*. In what follows, we will generically refer to $Acc(i)$ to refer to one of the two metrics presented above, when it will not be relevant to distinguish them.

*Time-expanded graph PT*

We need to compute travel times between any pair of centroids, to calculate areas $A(i,t_0)$, on which all the aforementioned accessibility metrics are based. To this aim, as in (18) and (23), we model PT as a *time-expanded graph*, which is the format underlying the General Transit Feed Specification (GTFS), the standard format describing PT scheduling. Each node of such a time-expanded graph is called *stoptime* and is characterized by a pair $(s,t)$, where $s \in \mathbb{R}^2$ is a physical PT stop (represented as a point in the plane) and $t \in R$ is a time instant in which a PT vehicle stops. So, for each PT line, a vehicle trip is modeled as a sequence of *stoptimes*. Let $ST\_l$ be the set of *stoptimes* of line l. Edge $(s,t)->(s',t')$, where $(s,t),(s',t') \in ST\_l$, models a vehicle trip within line l that starts at stop s at time t and arrives at stop s' at time t'. An edge $(s,t)->(s',t')$ between stoptimes of two different lines $(s,t) \in ST\_l$ and $(s',t') \in ST\_l'$ is added to indicate a transfer from a line to another if and only if (i) the distance between s and s' is no more than *max_walk_distance* and (ii) if it is possible to walk from stop $s$ at time $t$ and arrive at stop s' a time $t'' <= t'$. In this case, $t' - t''$ is the waiting time for the line $l'$. Let us denote by $G^{PT}$ the time-dependent graph constructed, for any line and all vehicle trips within each line, as explained.

Let us consider a user who with origin $x' \in \mathbb{R}^2$ and destination $x \in \mathbb{R}^2$, with a departure time $t_0$. There can be different ways in which the user journey can be performed. They can just walk, if $x$ and $x'$ are within *Dmax*. Or they can walk a short distance, reaching stop s, then board a PT vehicle at time $t$, i.e. *stoptime* $(s,t)$, and alight from the vehicle at stop $s'$, i.e. *stoptime* $(s',t')$: finally, they can walk from stop $s'$ reaching $x'$. The arrival time at his destination $x'$ is $t'$ plus the additional walking time. Transfers are also possible. We assume that users always choose, among the possible trip combinations, that one with the earliest arrival time. We perform user trip computations using CityChrone (22), without considering capacity constraints for PT vehicles.

**Our contribution: modelling DRT as a time-expanded graph and accessibility as a random variable**

To use the same methods and tools to compute accessibility used for conventional PT, our effort is to describe DRT with the same model, i.e., to extend the time-expanded graph of PT explained before with additional stoptimes and edges representing DRT. The main difficulty is that trips of DRT are not deterministic as described before for conventional PT. They can change from one day to another, to adapt to the different user requests. We therefore need to provide here a stochastic formulation.

Let us assume DRT operational characteristics have been fixed a-priori: there is a certain fleet of DRT vehicles, that all start from pre-fixed positions and a certain algorithm is applied to route DRT vehicles. Let $d$ be the stochastic process representing the user demand from 0:00 to 24:00. Let $d_\omega$ be one realization of such stochastic process, i.e., a sequence of user requests over 24h, each characterized by the time of request, the requested pick-up time, the origin and the destination. Observe that realizations $d_\omega$ are indexed by sample $\omega$ in some sample space $\Omega$. Sample $\omega$ can be interpreted as a day of DRT operation. We assume requests are generated from a same underlying stochastic process $d$, in the sense that realizations $d_\omega$ have all the same statistical properties, in terms of expected request rate and probability distribution of origins and destinations. Despite this statistical consistence, it is important to note that sequence $d_\omega$ of requests might be different from day to day.

Any sequence $d_\omega$ of user requests induces a certain set of vehicle trips performed by the DRT fleet over 24 hours. Let $P(d_\omega)$ be the function that returns the set of DRT vehicle trips over 24h to serve the requests in $d_\omega$. Each vehicle can perform multiple trips, one after another, with some inactivity time in between. We can represent vehicle trips in $P(d_\omega)$ as stoppoints and edges and add them to time-dependent graph $G^{PT}$ of conventional PT, in the following day. For each DRT vehicle, if it stops at location $x$ at time $t$ to pick up or drop off passengers, we add stoptime *(x,t)*. We then add edges between consecutive stoptimes covered by each vehicle. We add all these stoptimes and edges to $G^{PT}$ and we obtain new graph $G_\omega$. On this graph, we can compute, for any hexagon $i$, its accessibility $Acc_\omega(i)$. Note that, if we observe DRT operation on another day $\omega'$, we can see a different sequence of requests $d_{\omega'}$, and thus a different set of DRT vehicle trips $P(d_{\omega'})$ and thus a different resulting time-dependent graph $G_{\omega'}$. As a consequence, the value of accessibility $Acc_{\omega'}(i)$ in hexagon $i$ can also be different. In other words, for any hexagon $i$, its accessibility $Acc(i)$ is a random variable whose realizations are $Acc_\omega(i), \omega \in \Omega$.

Let $\mathbb{E}Acc(i)$ and $\sigma_{Acc(i)}$ be the expected value and the standard deviation, respectively. Quantity $\mathbb{E}Acc(i)$ indicates the accessibility that a user *can expect* at hexagon $i$ from the multimodal system composed of conventional PT plus DRT. Quantity $\sigma_{Acc(i)}$ expresses instead how *variable* such accessibility is. If $\sigma_{Acc(i)}$ is very small, it means that every day one can expect the same accessibility at hexagon $i$, i.e., everyday one can reach the same number of opportunities per hour.

Note that to compute $\mathbb{E}Acc(i)$ and $\sigma_{Acc(i)}$ one would need to observe an infinite number of days of operations, which is obviously impossible. We must thus be content with their estimations: if DRT operation has been observed a certain finite set $\widetilde{\Omega} \subset \Omega$ of days, we compute the empirical average $\widehat{Acc}(i)$ and the empirical standard deviation $\tilde{\sigma}_{Acc(i)}$ just on realizations $\{Acc_\omega(i) | \omega \in \widetilde{\Omega}\}$ to approximate $\mathbb{E}Acc(i)$ and $\sigma_{Acc(i)}$, respectively.

As we specify in the "Accessibility" section, we exclude from the analysis hexagons whose centroid is far away from any stoptime (further than *max_walk_distance*). It might thus happen that a certain hexagon $i$ is excluded from accessibility computation in some realizations $\omega$. This happens when neither conventional PT stops, nor any pickups/drop-offs of DRT are close to $i$ (it may be that no drop-offs and pick-ups in $d_\omega$ are inside $i$ or close enough, or it may also be that such requests might have existed, but the DRT service was unable to serve them). To make our analysis robust, we only calculate accessibilities of the hexagons that are never excluded in the observed days $\omega \in \widetilde{\Omega}$. Furthermore, we provide values of accessibilities only for the hexagons with small enough standard deviation, $\tilde{\sigma}_{Acc(i)} \leq \sigma_{max}$, as otherwise the measure would be unreliable.

**IMPLEMENTATION**

The implementation of the method is composed of five steps, as in Figure 1:

1) Generation of the demand for multiple days of observation, i.e., set $\{d_\omega | \omega \in \Omega'\}$, where $d_\omega$ is the sequence of user trip requests over one day of observation

2) Simulation, in Visum (24), of the operation of DRT to serve $d_\omega$ and store the resulting DRT vehicle trips $P(d_\omega)$, for $\omega \in \Omega'$.

3) Generation, via appropriate Python scripts, of time-expanded graph $G_\omega$ modeling DRT and conventional PT and store this graph in GTFS format, for each day of observation $\omega \in \Omega'$.

4) Computation of $Acc_\omega(i)$ using CityChrone (22), for all hexagons $i$ and for all days of observation $\omega \in \Omega'$.

5) Computation of average accessibility $\widetilde{Acc}(i)$ and standard deviation $\tilde{\sigma}_\omega(i)$ for all hexagons $i$, using Python and Excel.

Note that steps 1 and 2 are not needed when DRT trip observations come from a real system.

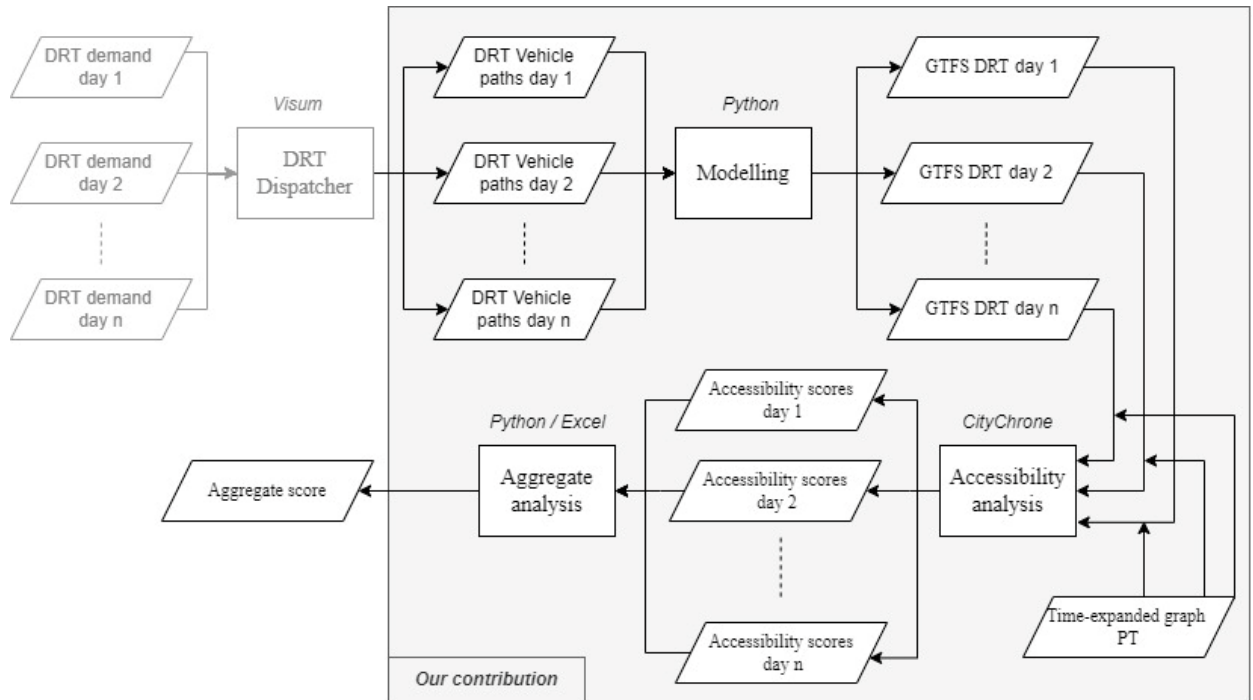

*Figure 1 Implementation procedure*

**Generation of demand $d_\omega$ and vehicle trips $P(d_\omega)$ in Visum**

First, we identify the sub-network where DRT service can be available, i.e. selected links where DRT vehicles are authorized to run; then we identified physical or virtual stop points for passengers Pick-Up and Drop-Off from the vehicle, named PUDOs.

To generate a demand $d_\omega$ for different observation days $\omega \in \Omega'$, we start from an hourly Origin-Destination (OD) matrix, containing the number of users willing to go from an origin zone to a destination zone, via DRT, at each hour. Each user is converted into a request in $d_\omega$, characterised by an origin point and destination point and a departure time. These two points are uniformly distributed at random within the respective origin and destination zones. Similarly, the departure time is also uniformly distributed at random in the corresponding 1h interval. We then simulate the operation of a fleet of DRT vehicles to serve demand $d_\omega$, using the dispatching algorithm of Visum (25) (but the procedure is not impacted by the dispatching algorithm employed).

Note that our procedure is oblivious to the operational rules and constraints of the DRT system under study. It can indeed be applied to many-to-one or many-to-many systems, stop-based or door-to-door, with prebooking or not. As long as a set of observations $P(d_\omega), \omega \in \Omega$ are available, our analysis can be performed, no matter the characteristics of the DRT service that generated such observations.

**Generation of time-dependent graph $G_\omega$ and GTFS representation**

From the node-level information contained in output $P(d_\omega)$ of the simulation, we create a list in which each row represents the position of a DRT vehicle, with the following attributes:

- *Vehicle id*
- *Node_ID,*
- *Node_coordinates,*
- *Is_it_stop*: indicating whether in this node the DRT has done a stop for a drop-off or pick-up
- *Arrival_time,*
- *Departure_time,*
- *Dwell time* (if provided) of a vehicle when the node is a stop;
- *Pass_on_board, Pass_boarded, Pass_alighted*: number of passengers on board, boarding and alighting at this node.
- *Prev_section_length*: travelled distance along the network between the current node and the previous one indicated within the previous row

We perform the following operations:

1. We remove the rows with $Pass\_on\_board = 0$, as in those moments the corresponding bus is not providing a service to anyone;
2. We group the consecutive subsequence of nodes where $Pass\_on\_board > 0$ in a single vehicle *trip,* and assign a *Trip_ID.*

The set of such vehicle trips constitutes $P(d_\omega)$, which is stored in the form of two files: *stop_times*.txt, with the list of stops with their associated transit times and *trips.txt* with all performed vehicle trips.

In addition, we also consider a third file with the list of service-enabled stops (PUDO). All this information is converted in the GTFS format and merged with the GTFS files of conventional PT. The result the time-dependent graph $G_\omega$.

We repeat the procedure above for all the considered days of DRT operation.

**Computation of accessibility $Acc_\omega(i)$ in CityChrone**

To calculate $Acc_\omega(i)$ for each day $\omega$ of DRT operation, we use the open-source platform CityChrone. The input data needed for the accessibility analysis are a shape file with the number of opportunities in each hexagon and the GTFS describing time-dependent graph $G_\omega$. Within CityChrone, the study area is tessellated in hexagons, and the number of opportunities contained in the aforementioned shape file is distributed in the hexagons in the corresponding locations. Hexagons can be arbitrarily small, however increasing the computation time.

The calculation of pedestrian routes between centroids of hexagons and stops is performed via the open-source routing engine, i.e. Open Source Routing Machine (OSRM). At this stage, we can identify the number of hexagons "served by public transport", which represent the service coverage. As output, for each day of operation $\omega \in \Omega'$, CityChrone provides accessibility scores $Acc_\omega(i)$ of all centroids $i$, except those whose centroid is not within distance *max_walk_distance* from a stoptime in $G_\omega$ (either of conventional PT or DRT).

Finally, we aggregate the results of the different days of operation $\omega \in \Omega'$, in Python, to compute, in each hexagon $i$, the empirical average $\widehat{Acc}(i)$ and the empirical standard deviation $\tilde{\sigma}_{Acc(i)}$.

## EVALUATION

### Description of the use case

We test the methodology in a case study to evaluate the accessibility that would result from a hypothetical implementation of a many-to-many DRT service, integrated with conventional PT, within the municipality of Acireale (Italy) (Figure 2).

Acireale is part of the metropolitan city of Catania, located in Sicily (Italy). It is characterized by a low frequency and poor coverage bus service, unable to provide a good connection with the main intermodal hubs (i.e. intercity bus terminals located in Piazza Livatino and Piazza Pennisi and train stations of Acireale and Guardia Mangano). The demand for PT is weak and scattered, making the implementation of DRT potentially appropriate. It is important to note that Acireale has a strong interaction with Catania, the main city in the metropolitan area (Figure 2-center), where most of the opportunities are. Figure 2-right depicts the sub-network of Acireale where the DRT service is allowed to operate, highlighting the location of main intermodal hubs and the *holding areas*, i.e., the locations where empty DRT vehicles can park, waiting for the next user trip request.

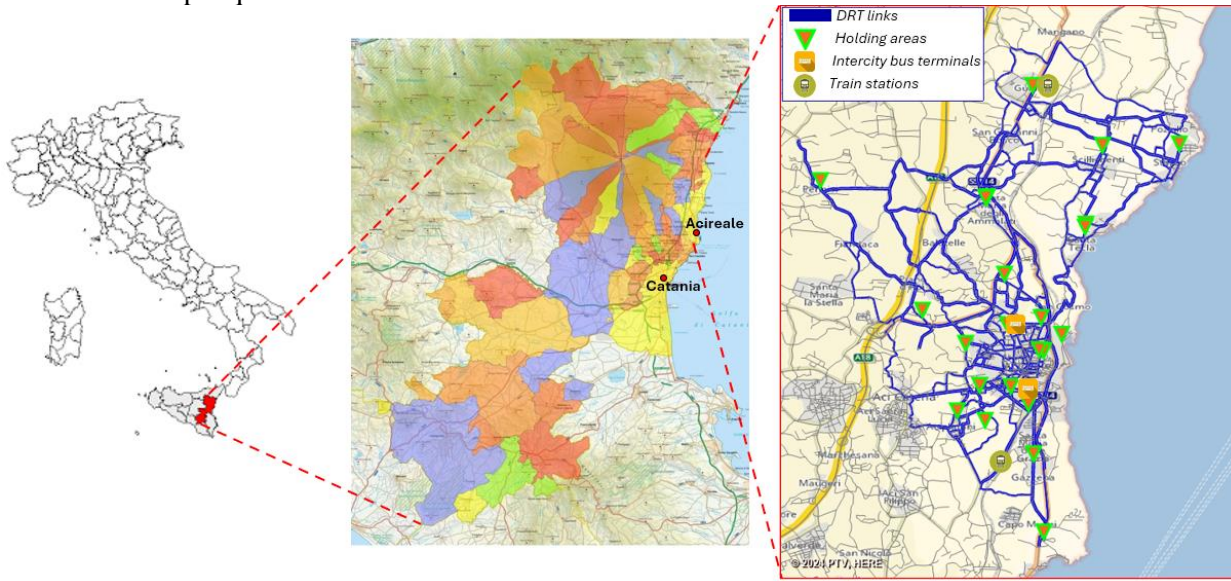

*Figure 2 DRT area and sub-network of the simulated DRT service within the municipality of Acireale*

Regarding the DRT demand, we consider the systematic movements provided by the Italian National Institute of Statistics (26); specifically, we start by considering the total number of users currently using PT. We further increase such number by adding the total percentage of users who travelled by car as passengers, as we assume that those passengers can benefit the most from improved PT. We then add to our demand 5% of users who currently drive their private vehicle, assuming they may shift to PT due to the improved performance brought by integrating DRT. Table 1 indicates the numbers of users served by DRT during an hourly simulation.

The simulated DRT service operates under the following time constraints, according to the dispatching policy implemented in VISUM (25) with the aim of minimizing the number of vehicles. For a specific user trip request, let us define the *Ideal Travel Time (ITT)* as the travel time of a vehicle hypothetically travelling along the direct distance from the origin to the destination. Let *DRTtime* be the actual in-vehicle travel time experienced by the user considering the detours. The difference between the *DRTtime* and *ITT* is the *Detour Time (DT)*. The *Detour Factor* of a user trip is *DF=DT/ITT*.

We also define the following parameters *Maximum detour factor (MaxDF)*
- *Maximum detour time (MaxDT)*

- *Detour time always accepted (AllAccDT)*

When a user issues a request, the dispatcher computes a feasible travel plan, trying first to use the vehicles already serving other users. The plan is proposed to the user, and we assume that:

 (i)  if $DT < AllAccDT$, the user accepts the plan;
 (ii)  if $DT > MaxDT$, the dispatcher assigns the user to a new empty vehicle if available;
 (iii)  if $AllAccDT < DT < MaxDT$ then.
    - if $DF < MaxDF$ the user accepts the plan; otherwise,
    - if $DF > MaxDF$, the dispatcher assigns the user to a new empty vehicle if available;

If there are no vehicles available for conditions (ii) and (iii), the request is rejected. Additionally, we also impose a maximum waiting time at the DRT stop (*MaxWait*) and a maximum walking time (*MaxWalk*) and the request will be rejected if these additional time constraints are not respected. We also fixed a maximum pre-booking time to make the request in advance, which could reduce the waiting time. We dimension the DRT fleet to obtain a service that is *predictable*, i.e., of which it is possible to have a good a-priori estimation of travel times. This property is important for user acceptance, in particular for DRT with no (or short) prebooking. Indeed, a user may need to know, with sufficient precision, how much time their trip is going to take, even if the trip will be requested hours or days later. With this information a user can determine if DRT is a convenient option for their trip. If it were not possible for a user to have a (at least approximate) idea of the travel time, they may perceive the DRT not reliable and may not choose it. We developed a *predictability indicator* in a previous paper (27), where we applied it in the same case study of this paper. Here, we dimension the fleet size so that the *predictability indicator* is at most 87%, which means that it is possible to provide to the user a sufficiently small range of values that contains the actual travel time 87% of times (27).

Table 1 summarizes all the DRT service parameters, as well as the parameters used for calculating accessibility metrics in CityChrone.

| DRT operational parameters | |
|---|---|
| *Sub-network links* | 2744 links |
| *Stops* | 899 stops |
| *MaxDF* | 3.5 |
| *MaxDT* | 15 minutes |
| *AllAccDT* | 8 minutes |
| *MaxWait* | 15 minutes |
| *MaxWalk* | 15 minutes |
| *Max pre-booking time* | 15 minutes |
| *Time interval of service simulation* | 7:00 a.m. – 8:00 a.m. |
| *Number of vehicles* | 16 vehicles |
| *Vehicle capacity* | 8 passengers |
| *Percentage of satisfied users* | 88% |
| *Users served* | 166 users |
| **Parameters related to Accessibility used in CityChrone** | |
| *Walk speed* | 3.9 km/h (28, 29) |
| *Max_walk_distance* | 1300 m (28, 29) |
| *Hexagon_side* | 0.5 km$^2$ |
| *Number of Hexagons* | 4987 |
| *Hexagon_number_ service areas* | 295 |
| *Geographical distribution of population* | EUROSTAT (30) |
| Departure time $t_0$ | 7:00 a.m. |
| *Maximum tolerated variability in accessibility $\sigma_{max}$* | 35% |
| **Characteristics of conventional PT within the metropolitan area of Catania** | |
| *Lines (extraurban and urban)* | 150 lines |
| *Vehicle Trips* | 469 |
| **Characteristics of conventional PT exclusively within the municipality of Acireale** | |
| *Lines (only urban)* | 4 |
| *Vehicle Trips* | 26 |
| *Removed lines in Scenario 3 Reduced PT+DRT* | 4 |

*Table 1 Case study parameters*

We compare the following three scenarios, where we vary the service in Acireale while leaving the PT of the rest of the metropolitan area:

- Scenario 1_*PTonly*: PT bus lines currently operated in Acireale (no DRT);
- Scenario 2_*PT+DRT*: PT bus lines currently operated plus the DRT service in Acireale;
- Scenario 3_*ReducedPT+DRT*: Current PT supply with lines partially replaced in the areas where the DRT is implemented, removing all lines operating exclusively operating within the municipality of Acireale (Table 1).

**DRT service variability**

Multiple simulations have been performed, by varying the random seed, to generate the demand for multiple days of observation, i.e. $\{d_\omega | \omega \in \Omega' = \{1, …, 39\}\}$, to capture the variability that characterize DRT. A total of 39 simulations are carried out. For each demand $d_\omega$ we obtained a set of DRT vehicle trips $P(d_\omega)$ used to carry out the 39 simulations. Note that the accessibility computation of scenario *PT only* does not depend on the demand, as the corresponding time-dependent graph just models the GTFS data of the conventional PT, which is oblivious to the demand.

Figure 3 shows the number of vehicle trips performed by the entire DRT fleet in each simulation. Each realization $P(d_\omega)$ of the set of vehicle trips in a day is used to enrich the time-dependent graph of conventional PT (as we explained in the Methodology section) in both the *PT+DRT* and *ReducedPT+DRT* scenarios.

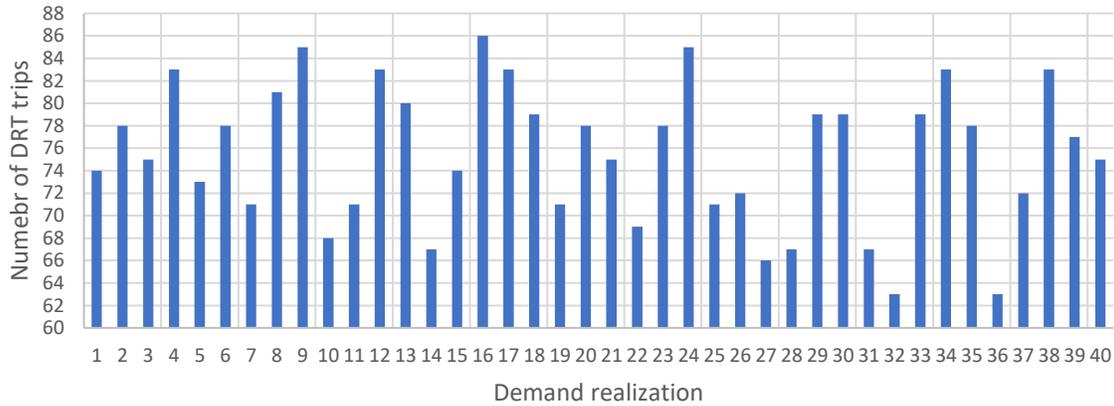

*Figure 3 Variability of simulated DRT service*

We calculate accessibility scores of the metropolitan city of Catania to consider the entire multimodal transport system and its interactions with the municipality of Acireale. However, we show the results of only the hexagons in Acireale, where the DRT service is implemented. Note that the opportunities (people) that are reachable from the hexagons in Acireale also include those in the rest of the metropolitan area.

We created the GTFS file (GTFS_base) of the metropolitan area by correcting the official GTFS of the PT operator in the area, affected by errors in the location of stops and scheduling. This GTFS is the implementation of graph $G^{PT}$. We compute accessibility scores on $G^{PT}$ for scenario *PT only*. The characteristics of conventional PT are in Table 1. We then remove the lines specified in Table 1 to obtain $G^{Reduced\,PT}$. Finally, we then build the GTFS representing the time-dependent graphs of *PT+DRT* scenario, by enriching $G^{PT}$ with the DRT trip vehicles. We similarly for Reduced PT +DRT, by enriching $G^{Reduced\,PT}$ this time.

Via CityChrone we obtain:

- For Scenario 1_*PT only*: 1 velocity scores and 1 sociality scores for each hexagon;

- For Scenario 2_*PT+DRT*: 39 velocity scores $V_\omega(i)$, $\omega \in \Omega' = \{1,..,39\}$ and 39 sociality scores $S_\omega(i)$, $\omega \in \Omega' = \{1,...,239\}$ for each hexagon $i$.
- For Scenario 3_*ReducedPT+DRT*: similar values as in Scenario 2_*PT+DRT*

For both scenarios *PT+DRT* and *Reduced PT + DRT*, we compute per each hexagon the empirical average $\widetilde{Acc}(i)$ and standard deviation $\tilde{\sigma}_\omega(i)$. We then compute the coefficient of variation as Equation 4:

$$CV(i) = \frac{\tilde{\sigma}_\omega(i)}{\widetilde{Acc}(i)} \cdot 100 \qquad (4)$$

Figure 4 shows the Empirical Cumulative Distribution Function (ECDF) of the coefficients of variation. To ensure that the accessibility metrics we will show in what follows are reliable, we exclude from the analysis all the hexagons whose coefficient of variation of the sociality score is larger than 35%- From Figure 4 one can see that this choice excludes only 20% of the hexagons in Acireale.

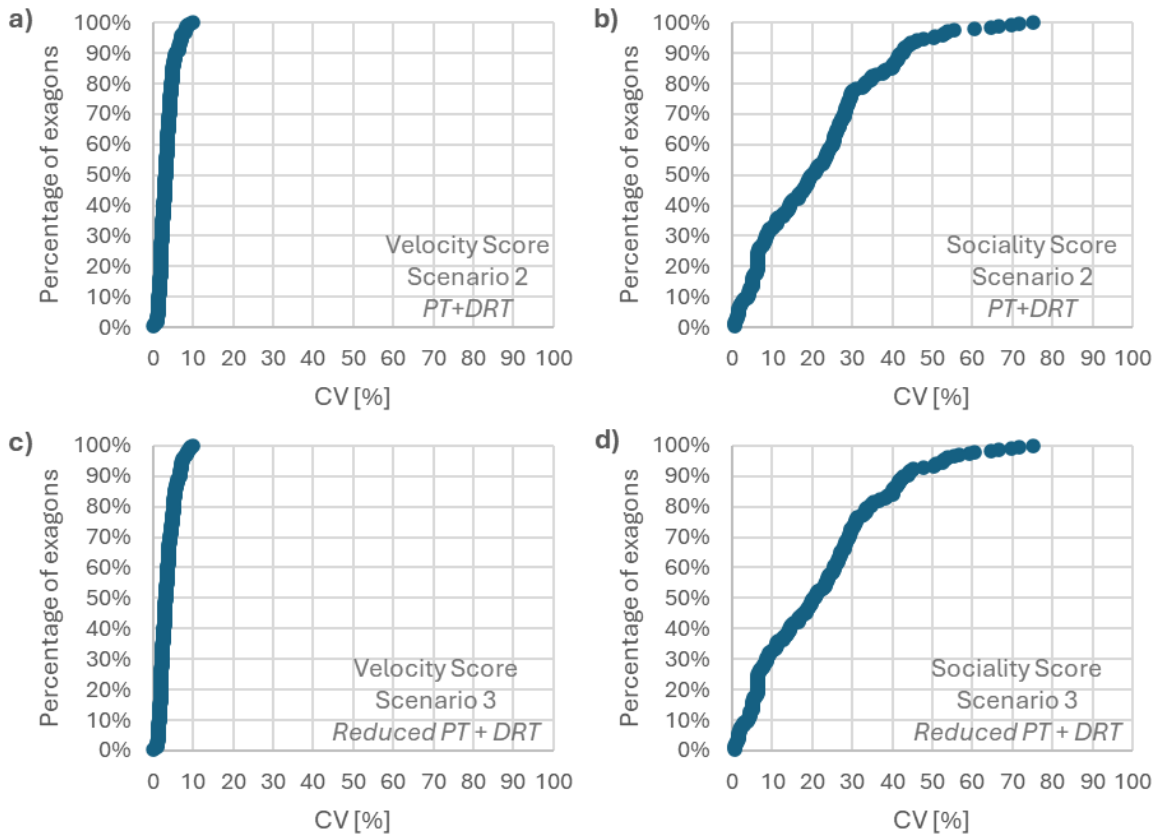

*Figure 4 ECDF of the coefficients of variation*

It is important to point out that in the 80% of the hexagons that we keep in our analysis, there are approximately 57,000 residents out of the total 60,000. In other words, we are still studying the accessibility experienced by 95% of the resident population.

**Improvements of accessibility**

We compare the calculated metrics in the three scenarios. ECDFs were used to correlate the accessibility scores and the percentage of population living in the analyzed hexagons.

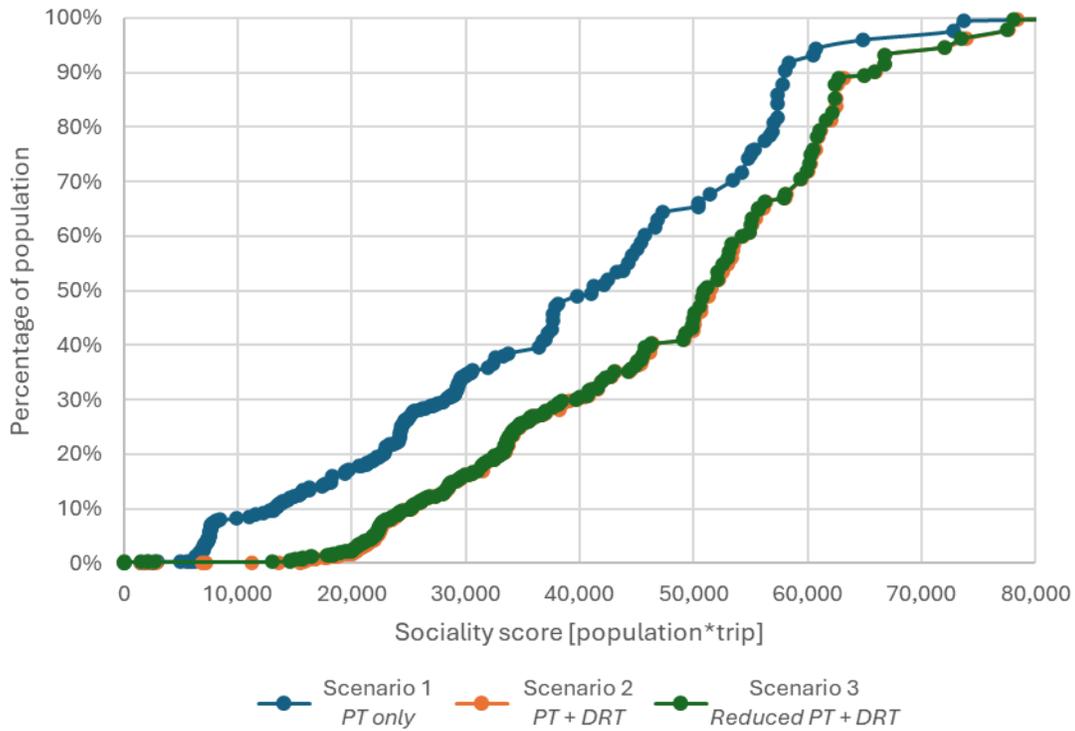

*Figure 5 ECDF of the sociality scores*

From the ECFDs in Figure 5 it emerges that the implementation of the DRT service improves accessibility, especially in the most disadvantaged hexagons (those with the lowest scores): The accessibility of the 10% population with the worst score jumps from ~12 000 people reachable per user trip to ~24 000 (100% increase).

The further remarkable point is that *Reduced PT + DRT* has no lower accessibility than PT + DRT: after integrating DRT, certain PT lines can be removed without hurting accessibility. The scatter plot of Figure 6 compares the sociality scores, hexagon by hexagon, of *PT only* and *Reduced PT+DRT*: for almost all hexagons accessibility increases, especially for the hexagons with very low values originally.

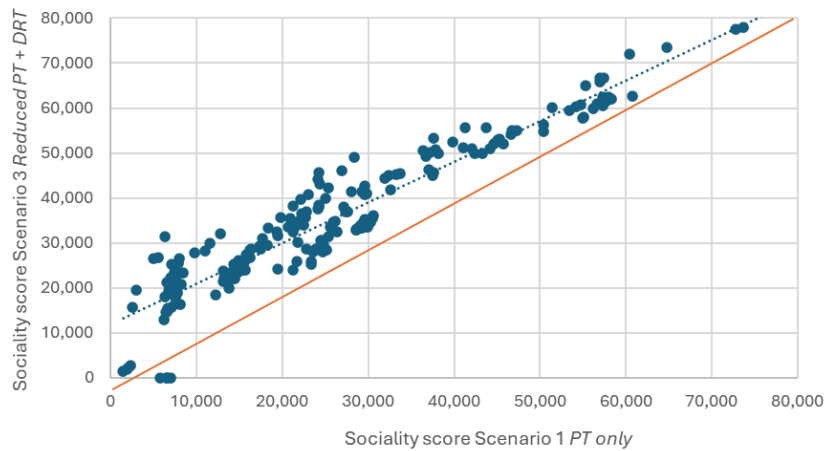

*Figure 6 Scatter plot of sociality score comparing Scenario 1 PT only and Scenario 3 Reduced PT + DRT*

Similar considerations emerge from the analyses conducted on velocity score (see Figure 7).

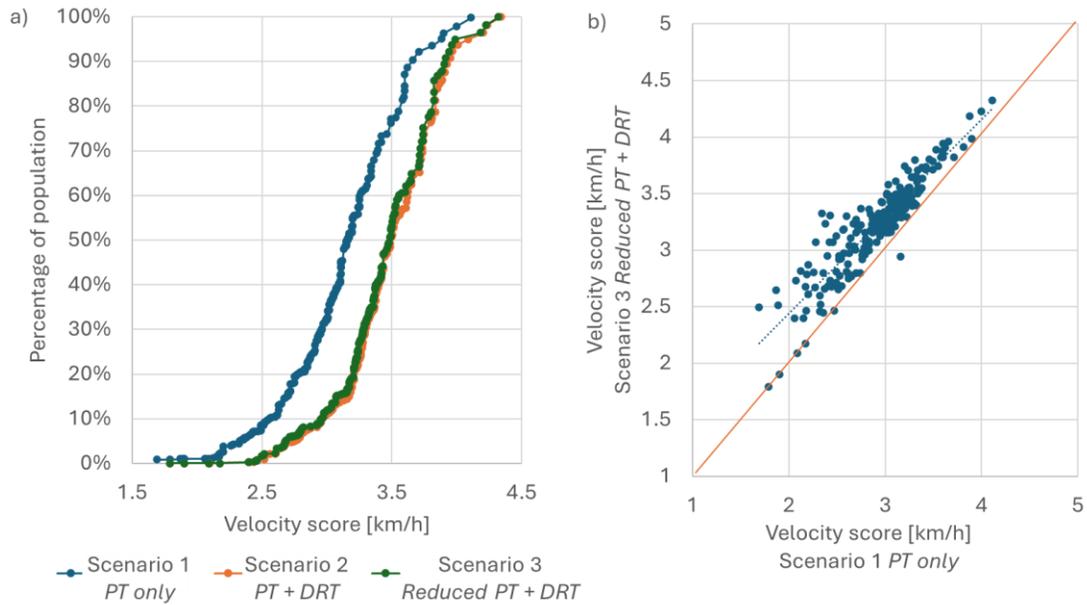

*Figure 7 a) Comparison of scenario velocity scores through ECDF, b) Scatter plot of velocity scores*

We now analyze the equality (horizontal equity) of the distribution of accessibility in Acireale. Using Lorentz curves (31), we can visualize and quantify equality both across hexagons and across the population Following the classic approach to construct Lorentz curves, to build Figure 8-left, in the x-axis we sort hexagons in ascending order of accessibility and we report in the *y*-axis the cumulative sociality score. In the figure, the *x* and *y* axis are normalized to 100%. To build Figure 8-right, as in (32), we assume that all residents in a certain hexagon enjoy the sociality score of that hexagon. Then, similarly as before, we sort residents in the *x*-axis in ascending order of accessibility, we report in the *y*-axis the cumulative sociality score and we finally normalize to 100% both axes. The dotted line corresponds to the ideal "perfect equality" situation, in which *X*% of hexagons (or population) would enjoy *X*% of sociality score. We observe that DRT reduces inequality considerably. Indeed, in the *PT only* scenario, 50% of hexagons only get 25% of sociality score, while in the Reduced PT+DRT, they get 35%.

Reduction of inequality across residents thanks to the DRT service is evident in Figure 8-right.

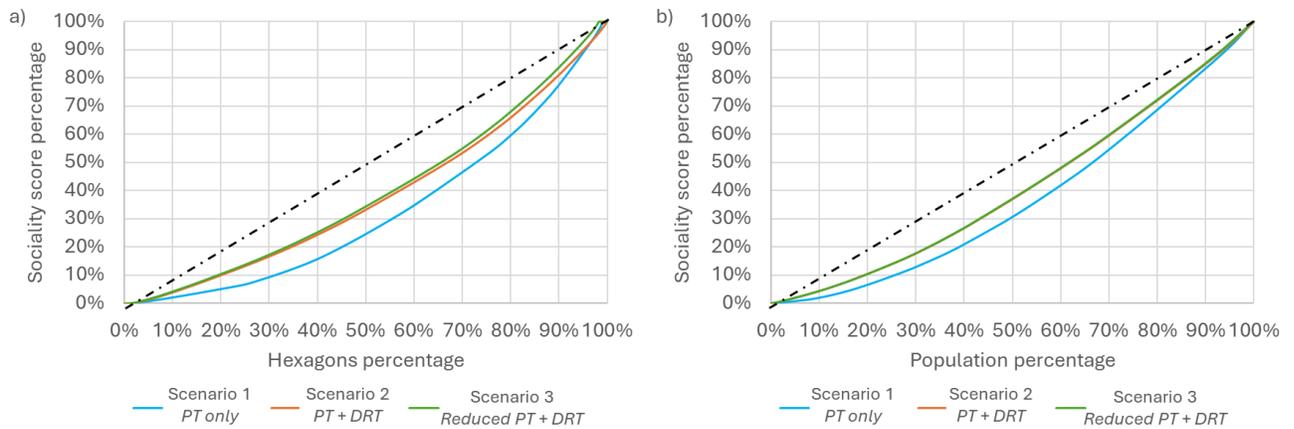

*Figure 8 Lorenz curve of sociality scores related to a) hexagons b) population*

The maps of Figure show *where* the improvement in accessibility concentrates. Since the simulated DRT is deployed in Acireale and its suburbs (Pennisi, Santa Maria, Ammalati, Scillichenti), this is where most of the improvement is. Improvement is especially concentrated in the abovementioned suburbans, in which PT supply only is insufficient (hexagons in dark and light grey in Figure 9-top-left become even blue in Figure 9-top-right). Albeit less evident, improvements in the accessibility score are also recorded in the central of Acireale. It is interesting to notice that some improvement of accessibility is also observed in nearby areas, where the simulated DRT does not operate, such as the neighbouring municipalities (i.e. Aci Sant'Antonio, Aci Catena and Aci S. Filippo), i.e., the number of hexagons in dark orange are observed increases from Figure 10-top-left to Figure 10-top-right. This means that people living there can now connect more easily to Acireale and reach the opportunities located therein.

Similarly, we realized the accessibility maps for the velocity score, shown in Figure 10. In this case, although the variations between Scenario 1 and Scenario 2 are less evident in absolute terms (i.e. increase in velocity values), the percentage variations registered between the various hexagons are in line with what was obtained for the sociality score.

Since the economic component cannot be ignored, we apply the methodology of *Associazione Trasportisti* (ASS.TRA) to estimate that the cost per passenger of the simulated DRT service would be 13% smaller than the cost per passenger with the current conventional PT. A more rigorous description of the cost estimation of simulated DRT can be found in (33).

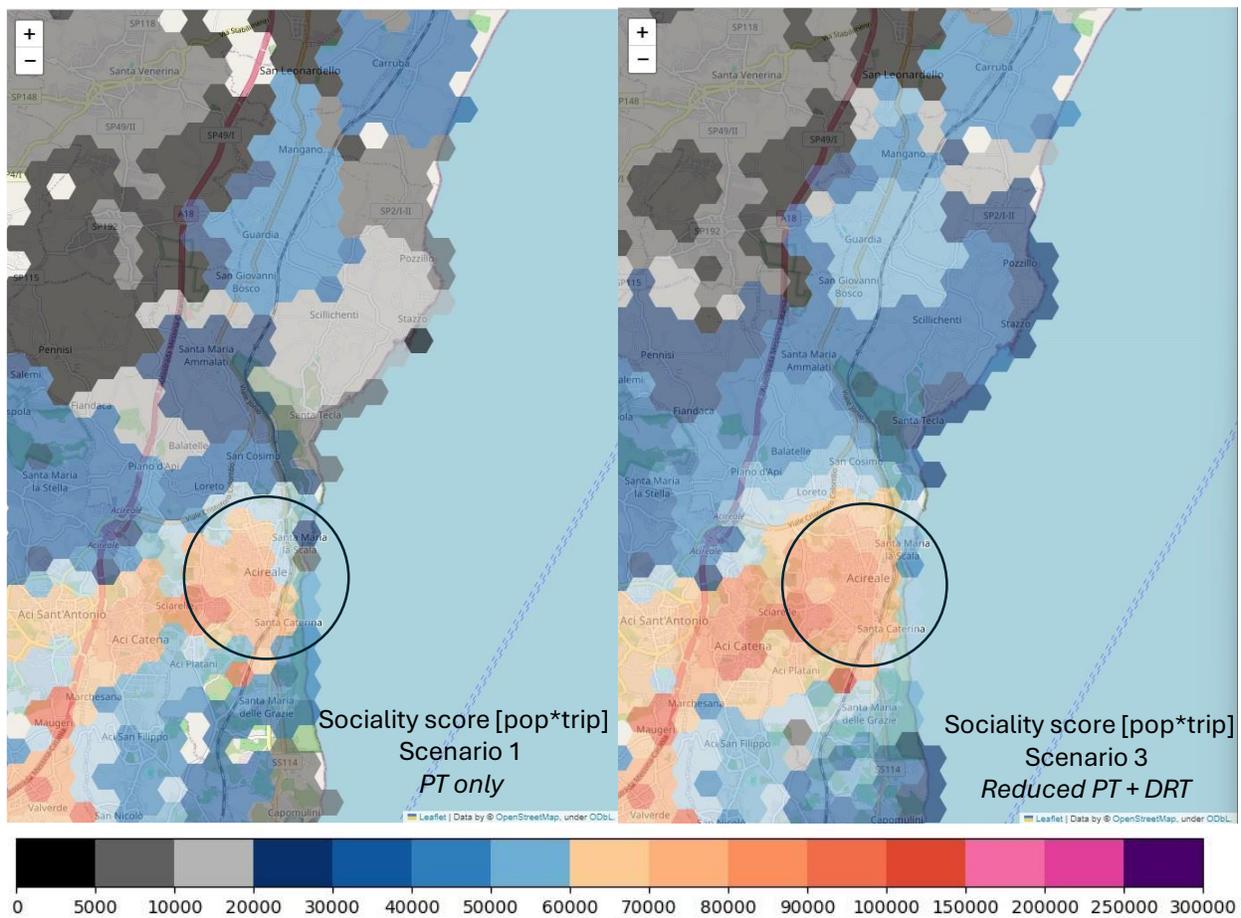

*Figure 9 Maps for Sociality score*

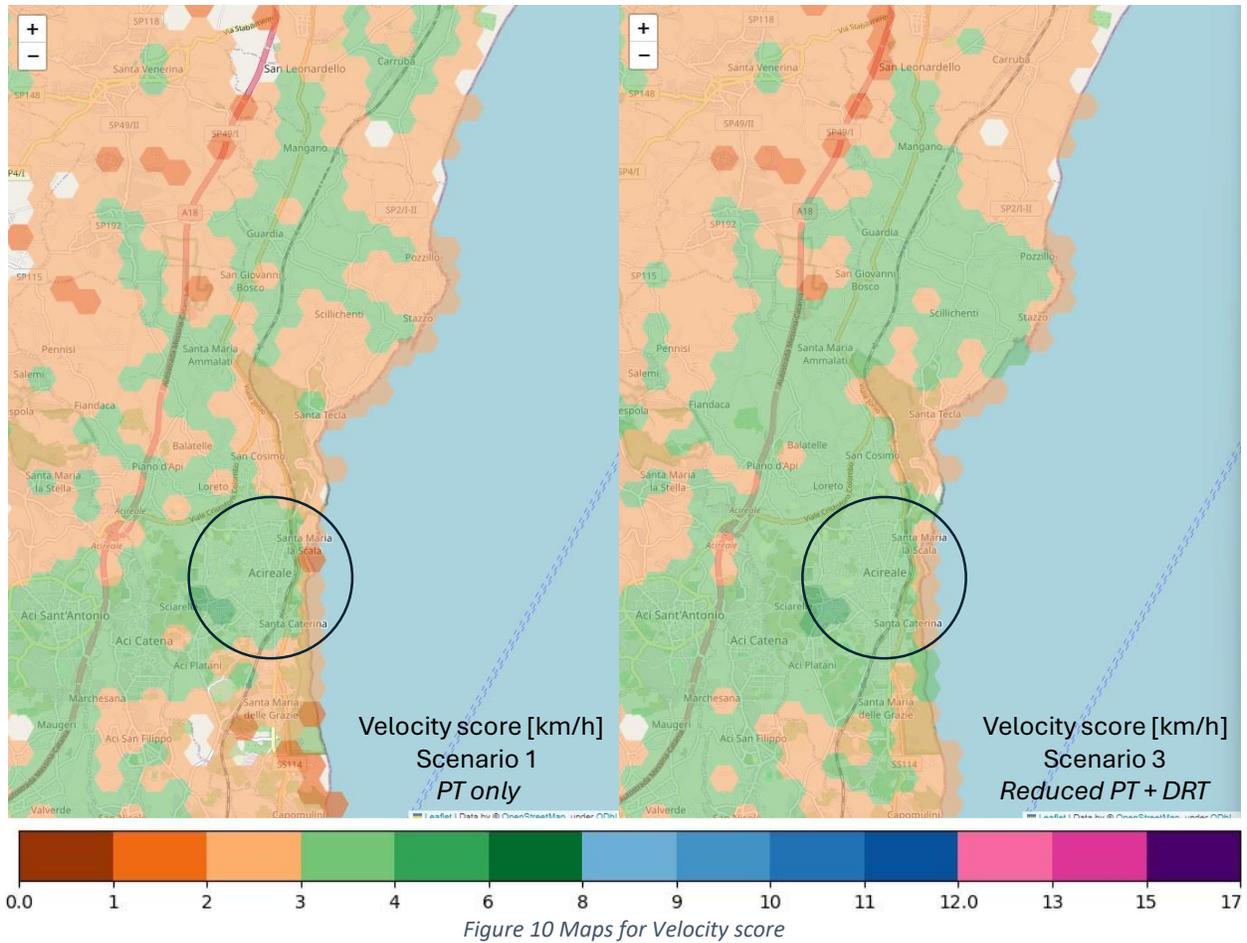

*Figure 10 Maps for Velocity score*

## CONCLUSIONS

Our study focuses on the integration of DRT systems within conventional PT and proposes a method to evaluate their impact on accessibility. The stochasticity of DRT makes classic methods for accessibility computation hard to apply. Our key idea to capture such stochasticity is to model the accessibility in each location as a random variable. To estimate such a random variable, we observe DRT over multiple days of operation, describe the operations of each day as a time-dependent graph, and compute accessibility on top of this graph. Finally, all the accessibility measures obtained over multiple days of operation are treated as the realizations of the "accessibility random variable". Our approach can be applied on observations from real-world pilots, or from simulation models, as in our case study.

We evaluated two accessibility scores, namely velocity and sociality scores, performing numerical and GIS spatial analysis, testing the methodology to a case study in Acireale. Our methods allowed to quantify the increase in accessibility achieved via integrating DRT in Acireale with the conventional PT of the broader metropolitan area. In our results, accessibility increased for 95% of the resident population in Acireale. The most significant improvements were registered when the population had a very poor transport service, where sociality score increases of over 90%. Consequently, (zonal) inequalities are strongly reduced, suggesting potential for reduction of car-dependency.

The richness of the findings we showcased in the use case of Acireale shows that our method can shed new light on the potential of DRT: while, up to now, DRT has been evaluated in terms of provided level of service, our method allows evaluate *accessibility*, which is a deeper a much more informative measure about the impact on a territory. We believe citizens and authorities will be more willing to invest in innovative services like DRT if they could quantify "*how many additional reachable opportunities per user trip such services can bring*". Our method enables this kind of analysis.

This work represents the core of a wider research effort that we are conducting, where we tackle service dimensioning, predictability, cost analysis and accessibility evaluation, to build a comprehensive methodology to support decisions of administrations and operators.

## ACKNOWLEDGMENTS

The work of Vincenza Torrisi is funded by the European Union (Next-Generation EU), through the MUR-PNRR project SAMOTHRACE (ECS00000022). This work has been supported by The French ANR research project MuTAS (ANR-21-CE22-0025-01).

## AUTHOR CONTRIBUTIONS

The authors contributed to the concept. P.L., A.A. and V.T. conceptualized the research objectives and developed the methodological framework. P.L. and V.T. developed the simulation model and P.L. designed the algorithm to conduct analysis. A.A. and V.T. provided advice and guidance throughout the process of algorithm design, analysis, and interpretation of results. A.A. and M.I. provided fundings. All authors reviewed the results and approved the final version of the manuscript.